\documentclass{PoS}
\usepackage{amsmath}

\title{Gravitational waves from cosmological first order phase transitions}

\ShortTitle{Gravitational waves}

\author{Mark Hindmarsh\\
  Department of Physics and Astronomy, University of Sussex,
  Falmer, Brighton BN1 9QH, U.K., and\\
  Department of Physics and Helsinki Institute of Physics, 
  P.O.Box 64, Fi-00014 University of Helsinki, Finland\\
\email{m.b.hindmarsh@sussex.ac.uk}}
\author{Stephan Huber\\
  Department of Physics and Astronomy, University of Sussex,
  Falmer, Brighton BN1 9QH, U.K.\\
\email{s.huber@sussex.ac.uk}}
\author{\speaker{Kari Rummukainen}\\
  Department of Physics and Helsinki Institute of Physics, 
  P.O.Box 64, Fi-00014 University of Helsinki, Finland\\
  E-mail: \email{kari.rummukainen@helsinki.fi}}
\author{David Weir\\
  Institute of Mathematics and Natural Sciences, University of Stavanger,
  4036 Stavanger, Norway \\
\email{david.weir@uis.no}}

\abstract{First order phase transitions in the early Universe generate
  gravitational waves, which may be observable in future space-based gravitational
  wave observatiories, e.g. the European eLISA satellite constellation.
  The gravitational waves provide an unprecedented direct view of the Universe 
  at the time of their creation.  We study the 
  generation of the gravitational 
  waves during a first order phase transition using large-scale simulations 
  of a model consisting of relativistic fluid and an order parameter field.
  We observe that the dominant source of gravitational waves is the sound
  generated by the transition, resulting in considerably stronger radiation than
  earlier calculations have indicated.
}

\FullConference{The 33rd International Symposium on Lattice Field Theory\\
		14 -18 July 2015\\
		Kobe International Conference Center, Kobe, Japan*}

\begin{document}

\section{Introduction}

The gravitational wave window to the Universe is about to open.  The
advanced LIGO gravitational wave laser interferometry detector
\cite{Harry:2010zz} has just started collecting data, and will soon be
joined by advanced VIRGO \cite{Accadia:2009zz} and KAGRA
\cite{Somiya:2011np} detectors.  These are expected to detect
gravitational wave signals from merging neutron star binaries and
possibly from supernovae.

In the early Universe there are several processes which may generate
observable gravitational radiation, such as inflation, cosmic strings
or other topological defects and first order phase transitions.  These
may be observable in proposed future space-based detectors, in the
first place the European eLISA satellite constellation \cite{Seoane:2013qna},
scheduled for launch in 2034.
It consists of three satellites in a triangular formation, which forms a laser
interferometer with arm length of order 1 million kilometers.  Due to its large size,
eLISA is sensitive to radiation at much lower frequencies than the Earth-based interferometers,
in the range $10^{-4}$--1 Hz.  eLISA technology demonstrator, LISA pathfinder, will be launched
in December 2015.

eLISA and other proposed space-based interferometers have the right frequency response for the detection of radiation form first order phase transitions at the electroweak scale and above.
In the Standard Model the electroweak transition is known to be a cross-over \cite{Kajantie:1996mn,Gurtler:1997hr,Csikor:1998eu,DOnofrio:2015mpa}, which does not lead to a gravitational wave signal.  However, a strong first order phase transition is possible in various extensions of the Standard Model~\cite{Carena:1996wj,Delepine:1996vn,Laine:1998qk,Grojean:2004xa,Huber:2000mg,Huber:2006wf,Dorsch:2013wja}.

A first order phase transition proceeds as follows \cite{Steinhardt:1981ct,KurkiSuonio:1984ba,Enqvist:1991xw}: due to the metastability associated with the first order phase transitions, the high-temperature ``symmetric'' phase supercools until  critical bubbles of the low-temperature ``broken'' phase are spontaneously nucleated.
These bubbles grow until the bubble walls collide with other bubbles and the phase transition is completed.  The growing bubble walls push the fluid along, causing hydrodynamical flows, which may persist long after the bubbles have vanished.

The generation of gravitational waves requires that the system has a non-vanishing quadrupole moment.  Thus, a single spherical bubble does not generate radiation.  However, when the bubbles collide the spherical symmetry is broken and gravitational radiation is possible.  In the 
widely used semi-analytical {\em envelope approximation} \cite{Kosowsky:1991ua,Kosowsky:1992vn,Kamionkowski:1993fg,Huber:2008hg,Caprini:2009fx} the fluid is modeled to behave like the order parameter field, and the gravitational waves are generated as the bubbles collide.  Thus, the gravitational waves originating from fluid flows after the transition has completed are ignored.

In this work we describe the phase transition using an effective relativistic fluid + scalar order parameter model, where the model parameters can be fixed to thermodynamic quantities of the original theory.  Using large-scale numerical simulations, we observe that the dominant source of the gravitational radiation are the acoustic waves generated by the bubbles: the sound of the transition. The role of the sound waves was originally suggested in ref.~\cite{1986MNRAS.218..629H}.
  Acoustic waves remain active long after the transition itself has completed, for up to the Hubble time.  The resulting gravitational radiation can be orders of magnitude stronger than indicated by  earlier results. 
 Our main results have been reported in \cite{Hindmarsh:2013xza,longbubbles}.

\section{Effective theory}

We describe the matter in the early universe  with a relativistic fluid coupled with a scalar order parameter field $\phi$ which drives the transition.  We note that $\phi$ needs not to correspond to any fundamental field of some underlying microscopic theory.  We use a potential of the form
\begin{equation}
V(\phi, T) = \frac{1}{2} \gamma (T^2-T_0^2) \phi^2 - \frac{1}{3} \alpha T \phi^3 + \frac{1}{4}\lambda\phi^4,
\end{equation}
where the parameters are adjusted to give the desired thermodynamic properties of the transition.
The rest-frame pressure $p$ and energy density $\epsilon$ are 
\begin{equation}
\epsilon = 3 a T^4 + V(\phi,T) - T\frac{\partial V}{\partial T}, ~~~~
p = a T^4 - V(\phi,T)
\end{equation}
with $a=(\pi^2/90)g$, where  $g$ is the effective number of relativistic degrees of freedom.    For details, we refer to \cite{longbubbles}.

The energy-momentum tensor of the field-fluid system is
\begin{equation}
\label{eq:tmunu}
T^{\mu\nu} = \partial^\mu \phi \partial^\nu \phi - {\textstyle \frac 12} g^{\mu\nu} (\partial\phi)^2
+ \left[\epsilon + p \right] U^\mu U^\nu + g^{\mu\nu} p
\equiv 
T^{\mu\nu}_{\rm field} + T^{\mu\nu}_{\rm fluid}
\end{equation}
where $U^\mu$ is the 4-velocity of the fluid.  The equations of motion are now obtained
from the energy-momentum conservation $\partial_\mu T^{\mu\nu}=0$, where we introduce
a non-unique coupling term permitting the transfer of energy 
and momentum between the field and the fluid \cite{Ignatius:1993qn,KurkiSuonio:1995vy,longbubbles}:
\begin{equation}
\partial_\mu T^{\mu\nu}_{\rm field} 
= -\partial_\mu T^{\mu\nu}_{\rm fluid} = \eta U^\mu\partial_\mu \phi \partial^\nu \phi.
\end{equation}
The final 
equations of motion can now be written as \cite{Hindmarsh:2013xza}
\begin{align}
  - \ddot{\phi} + \nabla^2 \phi - \frac{\partial V}{\partial \phi} &=
  {{\eta}} W (\dot{\phi} + v^i \partial_i \phi)  \label{eom1}
  \\
  \dot{E} + \partial_i (E v^i) + p [\dot{W} + \partial_i
  (W v^i)] - \frac{\partial V}{\partial \phi} W
  (\dot{\phi} + v^i \partial_i \phi) &=
  {\eta} W^2 (\dot{\phi} +
  v^i \partial_i \phi)^2
  \\
  \dot{Z}_i + \partial_j(Z_i v^j) + \partial_i p + \frac{\partial
    V}{\partial \phi} \partial_i \phi &=
  -{\eta} W (\dot{\phi} +
  v^j \partial_j \phi)\partial_i \phi.  \label{eom2}
\end{align}
Here $W$ is the relativistic $\gamma$-factor, $v^i = U^i/W$ the
fluid 3-velocity, $E=W\epsilon$ is the fluid energy density,
and $Z_i = W^2(\epsilon + p)v_i$ momentum density.
The right-hand sides of the equations (\ref{eom1}--\ref{eom2}) describe
the coupling of the field and the fluid, with the strength
parametrized by $\eta$.  

The traceless and transverse part of the energy-momentum tensor generates
metric perturbations:
\begin{equation}
      \ddot h_{ij} - \nabla^2 h_{ij} = 16 G T_{ij}^{TT}
\end{equation}
We use the procedure detailed in \cite{GarciaBellido:2007af} to project the
transverse part.

We refer to \cite{Hindmarsh:2013xza}  for the details of the lattice implementation of the equations of motion.
For the scalar field, we use the standard leapfrog update, and the the
relativistic fluid is evolved using the donor cell advection method \cite{WilsonMatthews}.
The lattice volumes vary up to $4200^3$, using up to 24\,000 cores on a Cray XC-30.
A somewhat different lattice implementation of the fluid+field system is presented
in ref.~\cite{Giblin:2014qia}.

\section{Results}

We show here results from simulations corresponding to relatively weak transition with latent heat ${\cal L} = (9/40) T_c^4$.  The phenomenological field-fluid coupling parameter is set to $\eta/T_c = 0.1$, $0.15$ and $0.2$.   For the detailed simulation parameters we refer to \cite{longbubbles}.  

\begin{figure}[t]
\vspace{-5mm}
\centerline{
  \includegraphics[width=0.25\textwidth]{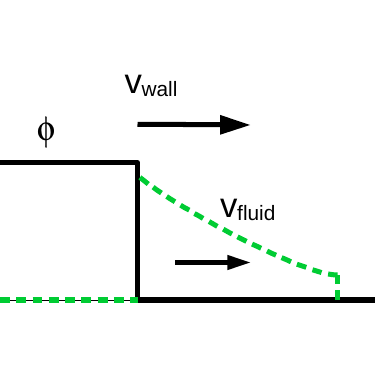}
  ~~~~~~~~~~~
  \includegraphics[width=0.25\textwidth]{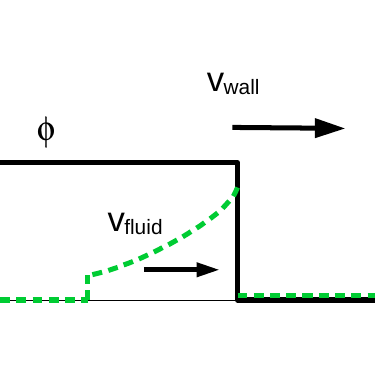}
}
\caption[a]{Deflagration (left) and detonation (right).  In deflagration
  $v_{\rm wall} < v_{\rm sound}$, and the growing bubble wall pushes the fluid 
  in front of it.  In detonation $v_{\rm wall} > v_{\rm sound}$, and 
  the bubble wall drags the fluid behind it.}
\label{fig:flows}
\end{figure}

When $\eta$ is small, the coupling between the field and the fluid is small, allowing the bubble wall to propagate quickly.  The moving bubble wall causes fluid flows.  The three values of $\eta$ are chosen so that we obtain three different bubble growth types:
at $\eta = 0.1$ the wall velocity is $v_{\rm wall} > v_{\rm sound} = 1/\sqrt{3}$ ({\em detonation}),
at $\eta = 0.15$ $v_{\rm wall} \approx v_{\rm sound}$ ({\em Jouguet}) and
at $\eta = 0.2$ $v_{\rm wall} < v_{\rm sound}$ ({\em deflagration}).  The moving bubble wall causes fluid flows: in deflagration, the wall pushes a thick layer (thickness $\propto$ bubble size) 
of fluid ahead of itself, whereas in detonation the bubble wall drags a layer of fluid behind it.
This is illustrated in Figure \ref{fig:flows}.


\begin{figure}[t]
\centerline{
\includegraphics[width=0.24\textwidth]{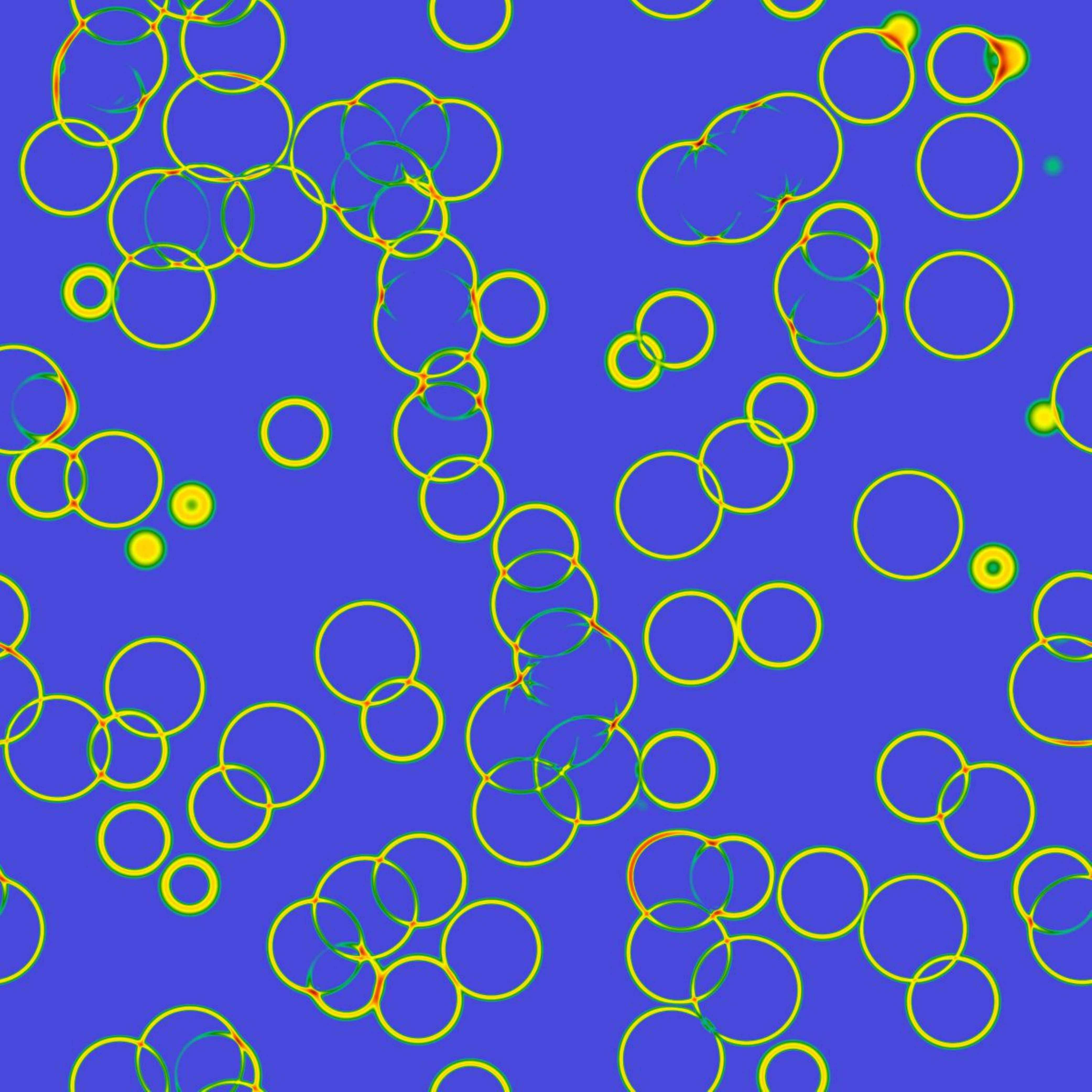} ~
\includegraphics[width=0.24\textwidth]{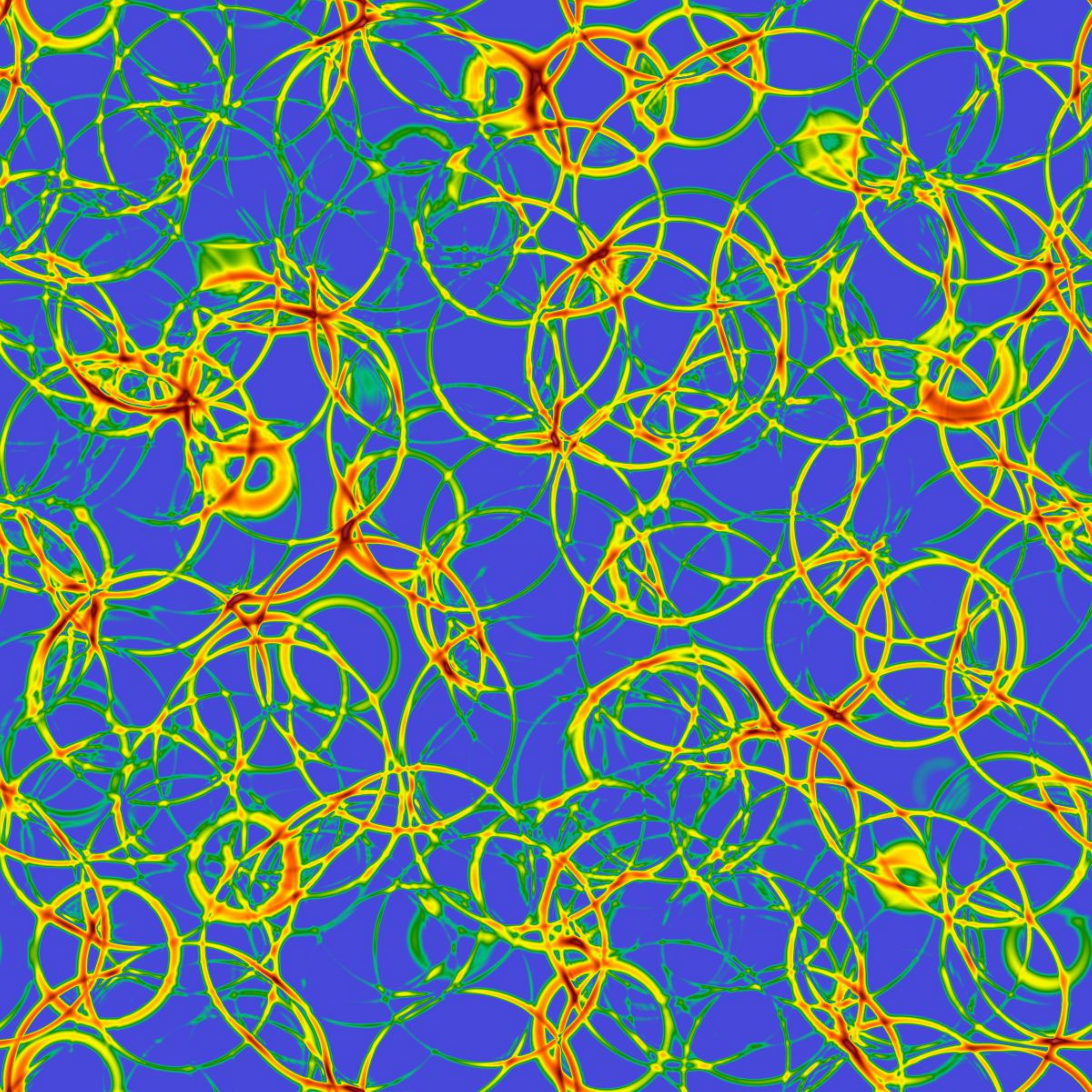} ~
\includegraphics[width=0.24\textwidth]{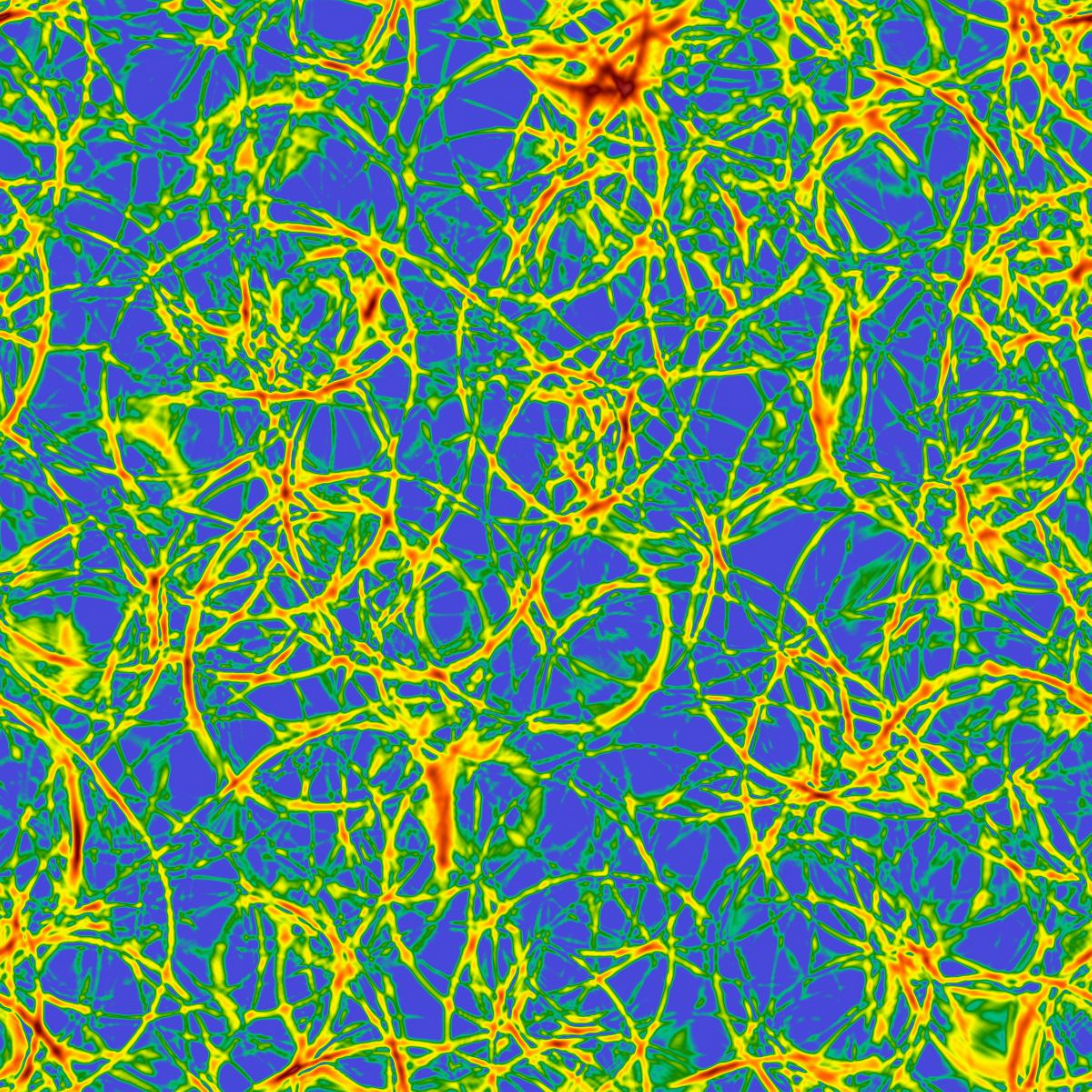} ~
\includegraphics[width=0.07\textwidth]{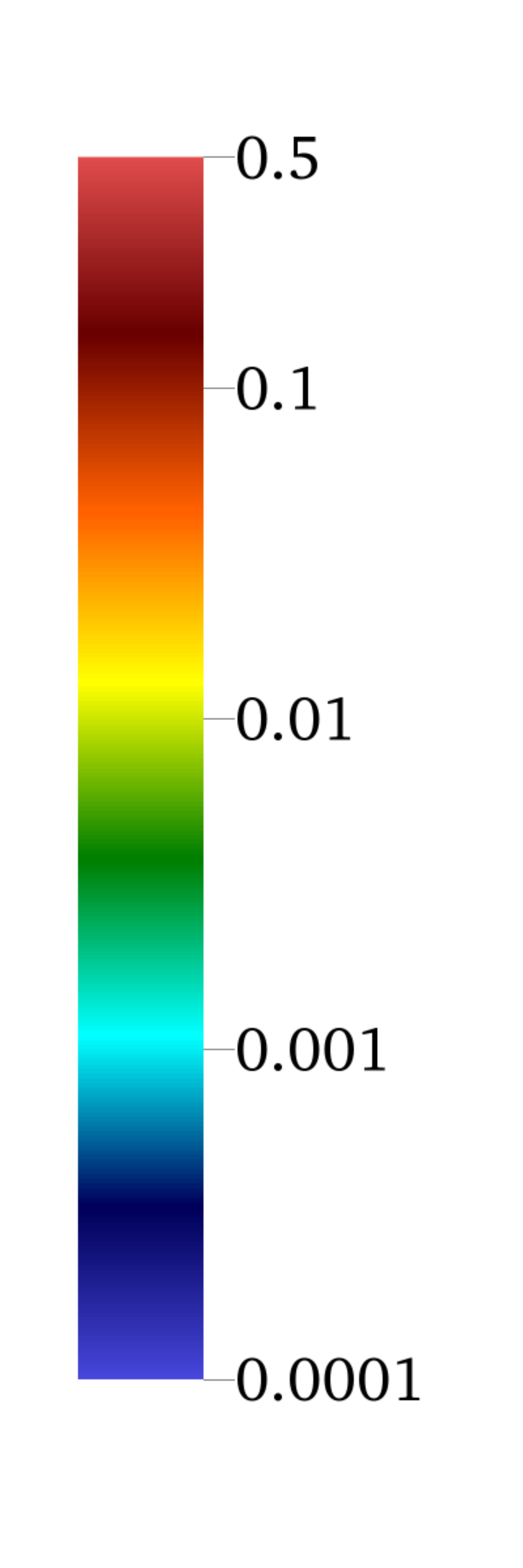}
}
\caption{Fluid kinetic energy density at $t=500/T_c$,
$1000/T_c$ and $1500/T_c$, at $\eta=0.15T_c$, corresponding to the growth phase of the bubbles,
end of bubble collisions and after the bubbles have vanished.  The shock waves caused by the bubbles remain for a long time after the transition has completed.}
\label{fig:bubbles}
\end{figure}

In Figure \ref{fig:bubbles}
we show three snapshots of fluid kinetic energy density 
from a simulation at $\eta=0.15$, taken at the bubble
growth stage, collision stage and after the bubbles have vanished.  During the growth
stage the kinetic energy is concentrated near the growing bubble walls.  After the bubbles
have collided the bubble walls vanish, but the fluid flow continues propagating as 
spherical compression waves, i.e. sound. 

\begin{figure}[t]
\centerline{
  \includegraphics[width=0.45\textwidth]{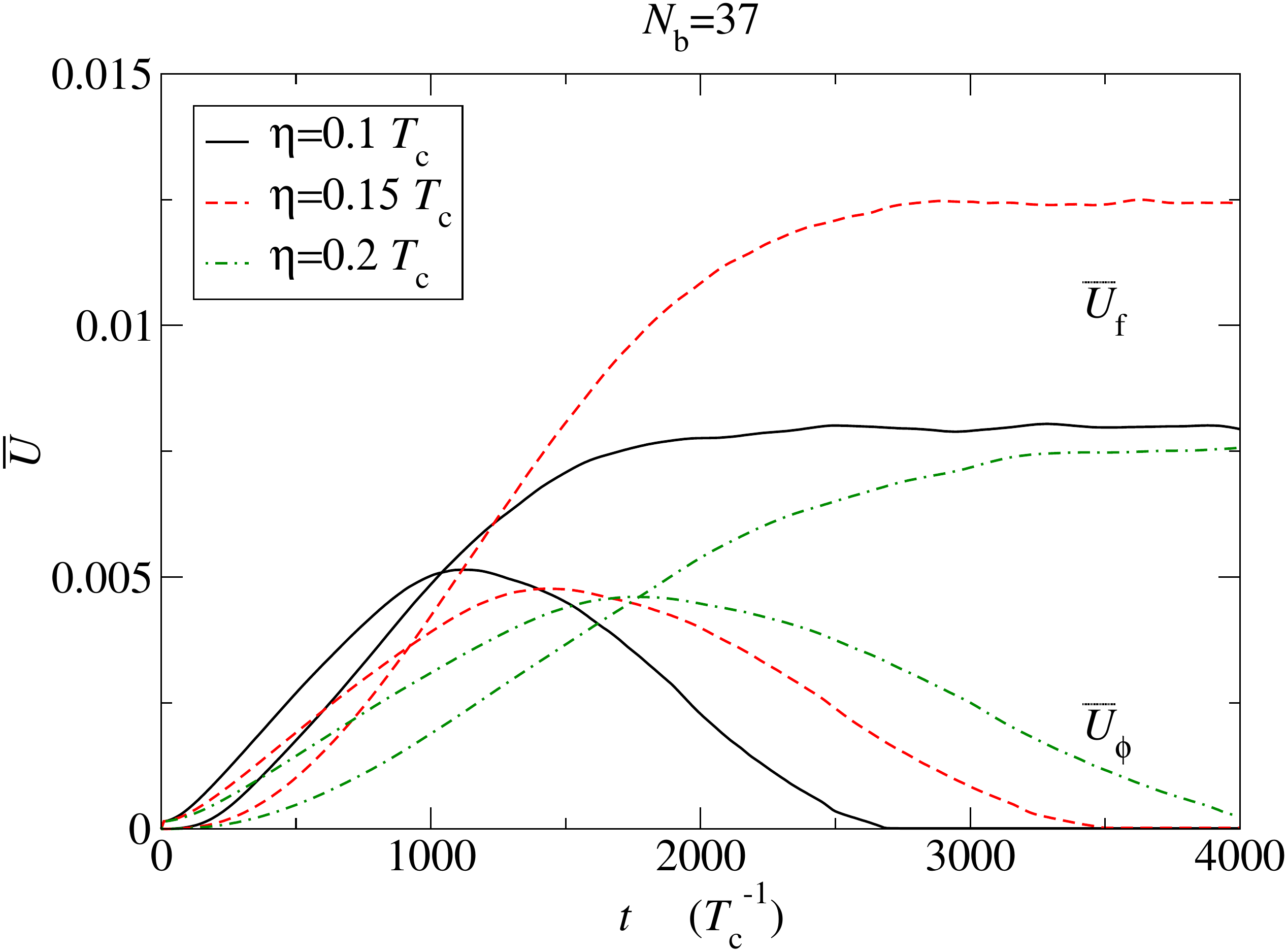} ~~~
  \includegraphics[width=0.45\textwidth]{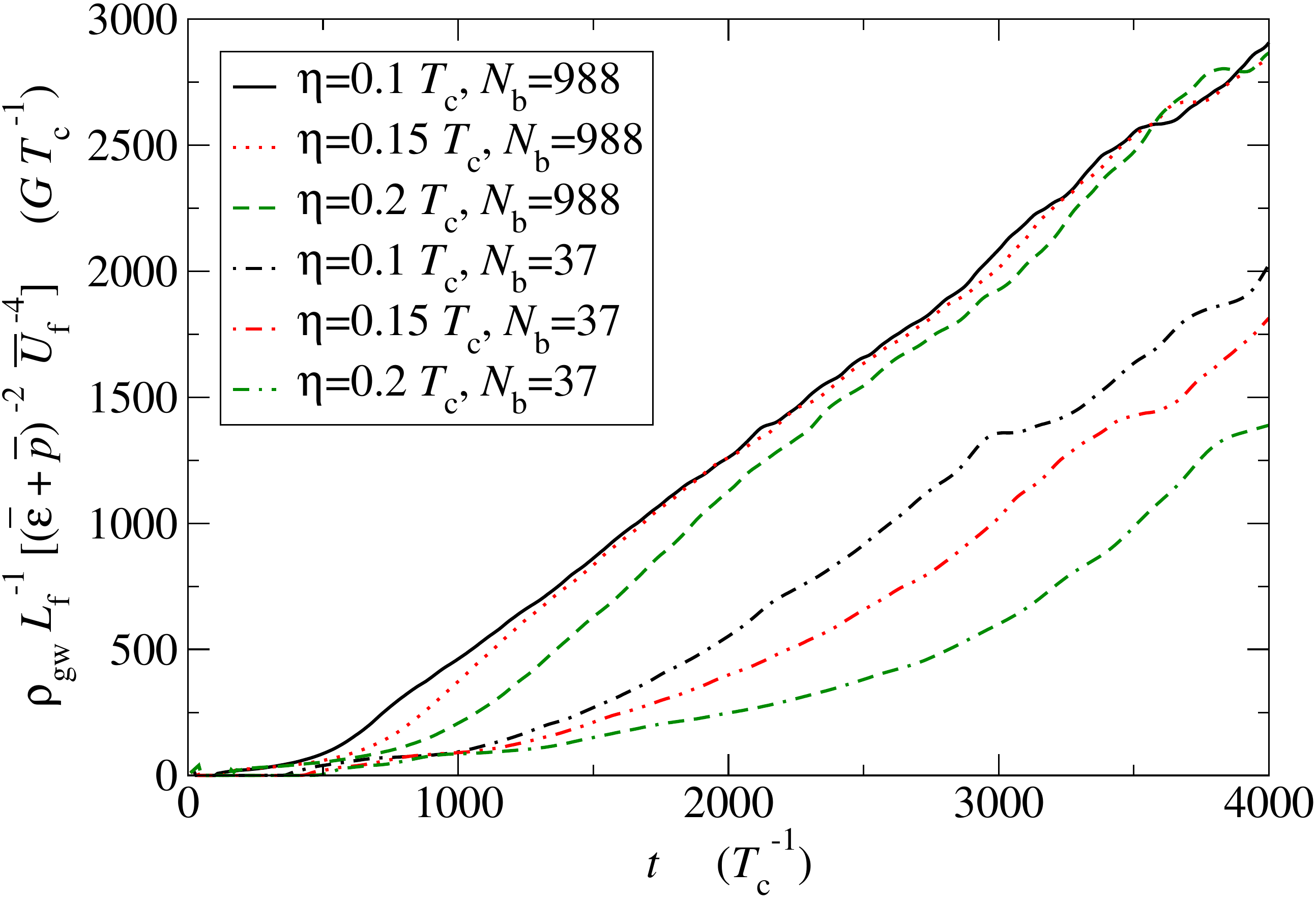}
}
\caption{Left: the relative contribution from the field ($\bar U_\phi$) and fluid ($\bar U_f$) to
  the energy-momentum tensor.  Right: the gravitational radiation power $\rho_{\rm gw}$ as a function of time.}
\label{fig:ubar}
\end{figure}

The contribution of the field and fluid to the stress-energy tensor (and hence gravitational
waves) can be quantified by introducing RMS fluid velocity $\bar U_f$ and the
equivalent field quantity:
\begin{equation}
        (\bar\epsilon + \bar p) \bar U_f^2 = \frac1V \int dV \tau_{ii}^{\rm fluid},
~~~~~~~~~~
        (\bar\epsilon + \bar p) \bar U_\phi^2 = \frac1V \int dV \tau_{ii}^{\rm field}.
\end{equation}
Here $\bar\epsilon$ and $\bar p$ are average energy density and pressure.
In Figure \ref{fig:ubar} we see that the field and fluid contributions are comparable only
during the bubble growth and collision stages.  After the bubbles have collided, $\bar U_\phi \approx 0$ but the fluid kinetic energy remains approximately constant.\footnote{%
The slow decrease in $\bar U_f$ 
is due to numerical viscosity of our simulation; the physical viscosity is negligible \cite{longbubbles}.}  
This implies that the gravitational wave production also remains active for much longer than the transition time itself; indeed, it can be estimated to continue for up to Hubble time \cite{longbubbles}.  From Figure \ref{fig:ubar} we can also see that the gravitational wave power grows linearly with an universal slope:
\begin{equation}
  \rho_{\rm GW} = t\,C\, G L_f (\bar\epsilon+\bar p)^2 \bar U_f^4,
  ~~~\mbox{with}~C = 0.8\pm 0.2
\end{equation}
Here $L_f$ is characteristic flow length scale.  We can estimate that the total power is up to two orders of magnitude stronger than the estimate from the envelope approximation.

\begin{figure}[t]
  \centerline{
    \includegraphics[width=0.45\textwidth,clip=true]{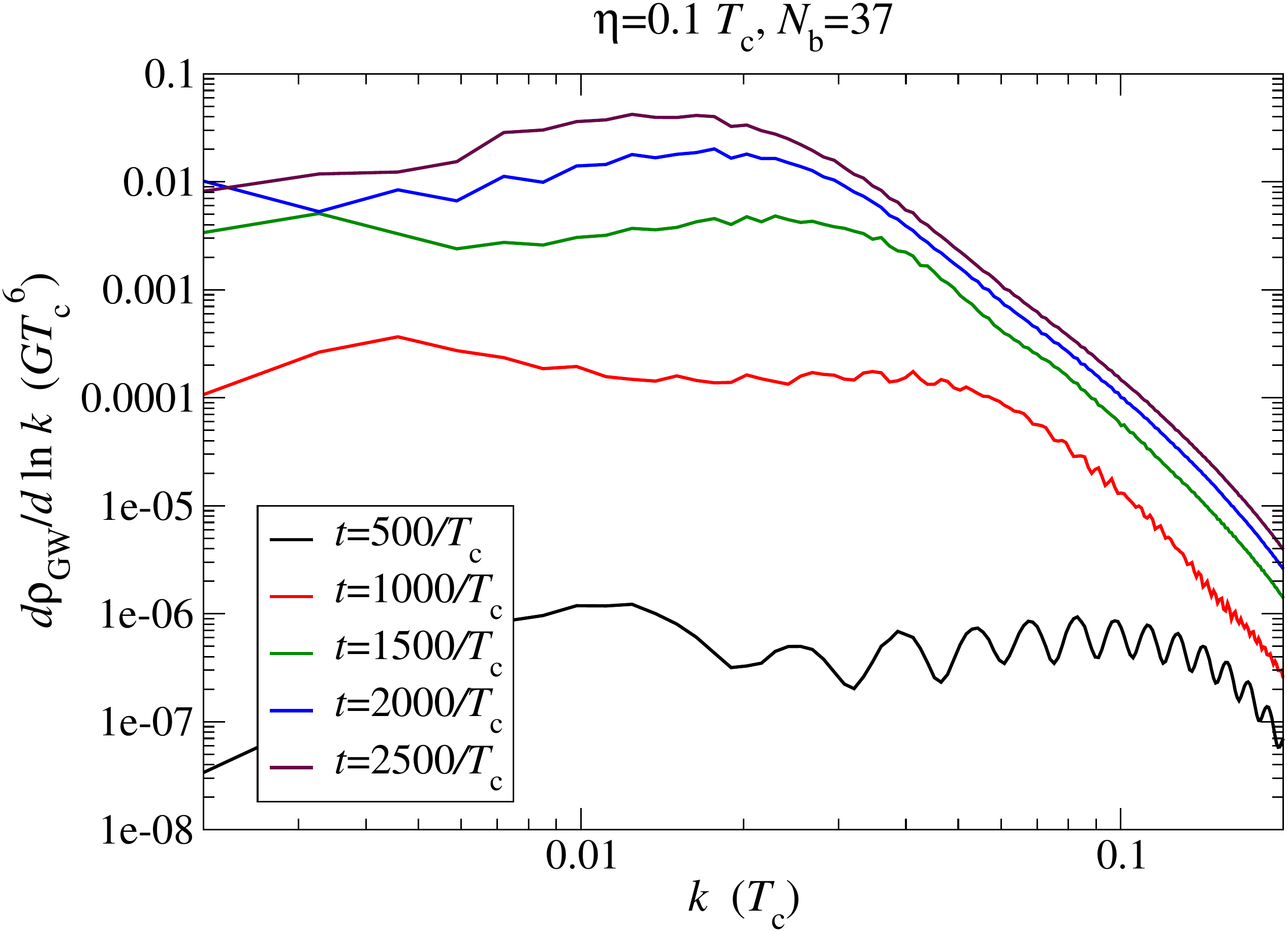}
    ~~~~~
    \includegraphics[width=0.45\textwidth,clip=true]{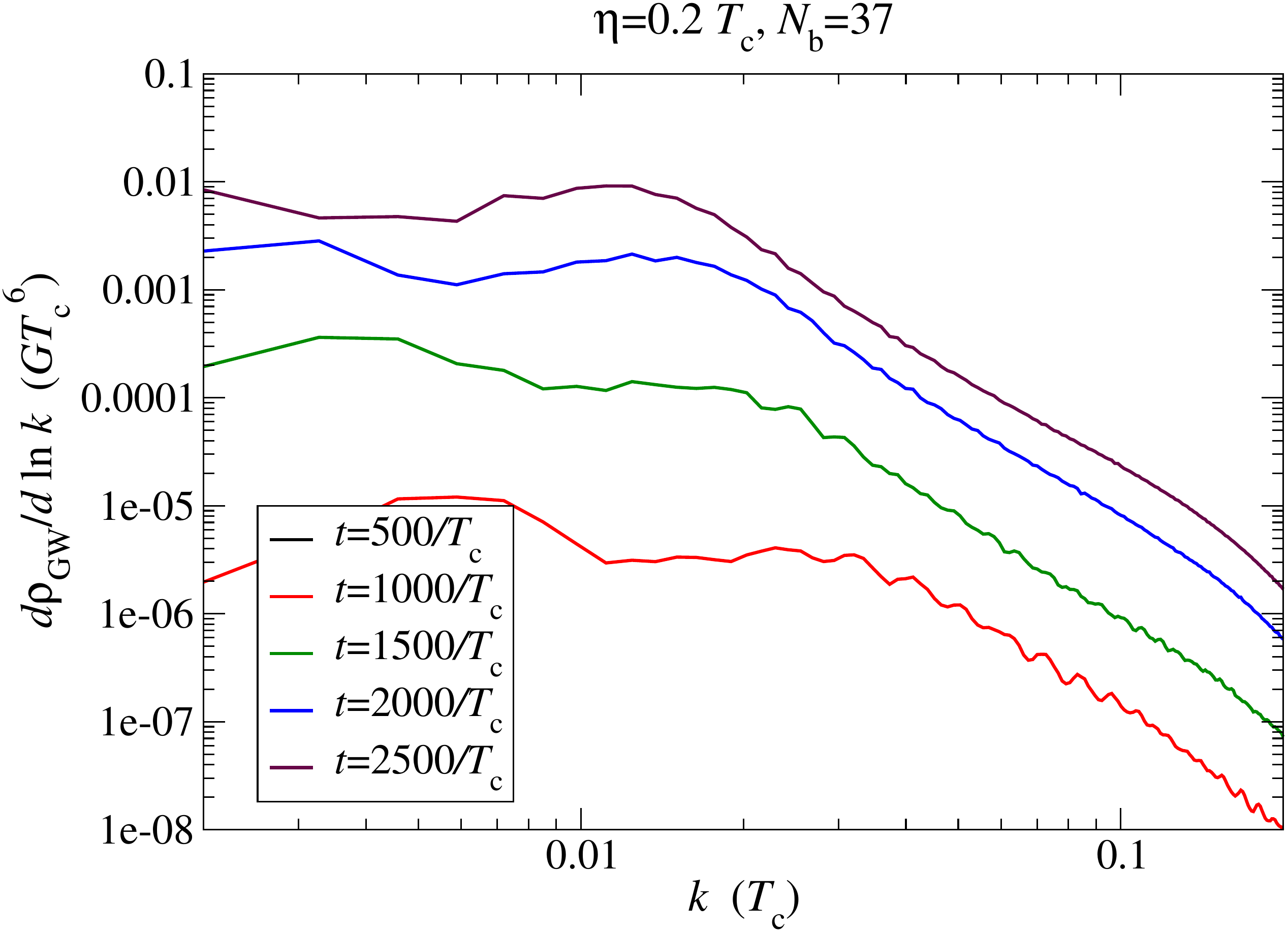}
  }
\caption{Gravitational wave power spectra from detonation (left) and deflagration (right).
  The development of characteristic power laws can be observed.}
\label{fig:spectrum}
\end{figure}

Finally, in Figure \ref{fig:spectrum} we show an example of the development of the 
gravitational wave power spectra from detonation and deflagration.  We see the development
of characteristic power laws, with possibly non-universal power.  For weak 
deflagration, we observe $\rho_{\rm GW} \propto k^{-3}$ on the UV end of the spectrum, deviating
strongly from the prediction from the envelope approximation ($\rho \propto k^{-1}$).

\section{Conclusions}

We have studied the production of the gravitational waves in first order phase
transitions in the early universe.  We observe that the dominant source of the
radiation is the {\em sound} of the transition, i.e. compression waves generated by the
growing bubbles.  These sound waves propagate through the universe long after
the transition has completed.  This strongly enhances the production of the
gravitational waves, giving up to two orders of magnitude stronger signal
than earlier eastimates.  This enhances the possibility of observation of 
primordial gravitational waves in future gravitational wave detectors.

\subsubsection*{Acknowledgments}
We have used the facilities at the Finnish Centre for Scientific
Computing CSC, and the COSMOS Consortium supercomputer (within the
DiRAC Facility jointly funded by STFC and the Large Facilities Capital
Fund of BIS). KR acknowledges support from the Academy of Finland
project 1134018; MH and SH from the Science and Technology Facilities
Council (grant number ST/J000477/1). DJW is supported by the People
Programme (Marie Sk{\l}odowska-Curie actions) of the European Union
Seventh Framework Programme (FP7/2007-2013) under grant agreement
number PIEF-GA-2013-629425.


\bibliographystyle{JHEP}
\bibliography{gwaves}


\end{document}